\documentclass[runningheads]{llncs}
\usepackage[T1]{fontenc}
\usepackage{graphicx}
\usepackage{csquotes}

\usepackage{makecell}

\begin{document}
\title{Experimentation in Early-Stage Video Game Startups: Practices and Challenges\thanks{This is the authors' version of the manuscript accepted for publication in the Proceedings of 14th International Conference on Software Business (ICSOB), 2023. This manuscript version is made available under the CC-BY-NC-ND4.0 license. http://creativecommons.org/ licenses/by-nc-nd/4.0}} 
\titlerunning{Experimentation in Early-Stage Video Game Startups}
%

\author{Henry Edison\inst{1} \and Jorge Melegati\inst{2} \and Elizabeth Bjarnason\inst{3}}
%
%
\institute{Blekinge Institute of Technology, Karlskrona, Sweden \\ \email{henry.edison@bth.se} \and Free University of Bozen-Bolzano, Bolzano, Italy \\ \email{JorgeAugusto.MelegatiGoncalves@unibz.it}
\and
Lund University, Lund, Sweden \\ \email{elizabeth.bjarnason@cs.lth.se}
}
%
\maketitle              
\begin{abstract}
Experimentation has been considered critical for successful software product and business development, including in video game startups. Video game startups need ``wow'' qualities that distinguish them from the competition. Thus, they need to continuously experiment to find these qualities before running out of time and resources. In this study, we aimed to explore how these companies perform experimentation. We interviewed four co-founders of video game startups. Our findings identify six practices, or scenarios, through which video game startups conduct experiments and challenges associated with these. The initial results could inform these startups about the possibilities and challenges and guide future research.
\end{abstract}

\section{Introduction}
	
\setcounter{footnote}{0}

Over the last 40 years, video games have increasingly replaced traditional games as leisure activities and have disrupted how we spend our leisure time. The video game market has become an established and ever-growing global industry for over two decades. In 2022, the global video market was worth USD 42.9 billion, and the revenue is expected to grow with an annual growth rate of 8.74\%\footnote{https://www.statista.com/statistics/292516/pc-online-game-market-value-worldwide/}. Originally, video games refer to the games that do not require a microprocessor and use analogue intensity signals displayed on a cathode ray tube (CRT) \cite{wolf08}. The availability of new imaging technologies, such as consoles, home computers, Virtual Reality (VR) devices, etc., has made the idea of video games more conceptual and less tied to a specific technology \cite{nielsen20}. 

Developing a successful video game is a very demanding and complex process. It involves expertise from various disciplines, e.g. software/game development, arts, animation, sound engineering, etc., which may increase the complexity of communication and coordination \cite{koutonen13}. Furthermore, it is unclear whether a game will succeed in the market, which poses a major risk to game publishers when investing in new game development projects. Unlike other software startups, video game startups do not build technological solutions to solve real problems. Instead, they combine art, science, and craft to offer fun, entertainment, and experience through the games \cite{crawford84,lewis11}. Yet, these requirements have no metrics to be applied, yet they must be validated at each stage of the development process. 

An effective adoption and implementation of experimentation is a staged process \cite{melegati20}. In this study, we aim to gain insights into how video game startups approach experimentation to develop games. To guide the study, we explore the research question: \textit{How do video game startups use experimentation in practice?} 

\section{Background and Related Work}
\label{sec:background}

In innovative endeavours, the required knowledge for success is generally unknown~\cite{kerr14}. Thus, experimentation is particularly useful for acquiring knowledge and reducing uncertainty. Experimentation is an approach based on continuously identifying critical assumptions, transforming them as hypotheses, and prioritising and testing them with experiments to support or refute them~\cite{Lindgren2016}. However, most startups persist with the original ideas rather than experimenting \cite{giardino14,gutbrod17,Pantiuchina2017}. 

While research in game startups exists, they are limited to mobile game development. For example,Vanhala et al.~\cite{Vanhala2014} analysed six Finnish mobile game startups and found that human capital is the most important element in their business models. Moreover, the key challenge is to raise the awareness of game players. Kasurinen et al. \cite{kasurinen13} showed that game developers are generally pleased by the tools available to experiment with the concept and build prototypes.

Research also shows that the iterative and incremental nature of agile methods positively impacts communication, game quality, and the ability to find the fun aspects of the mobile game features \cite{koutonen13}. In contrast, the agile principle of embracing changes increases the pressure to meet the deadline  \cite{borg20}. Mobile game startups should be cautious in considering the minimum viable product concept. The first version of a game artefact released to the market needs to be of sufficient quality to attract and lock in users for an adequate amount of time to allow for further development of the game \cite{roshan19}. This study aims to complement existing research by investigating how video game startups conduct experimentation.  

\section{Research Methodology}
\label{sec:rm}

We performed semi-structured interviews \cite{creswell09} to gain insights into how video game startups conduct experimentation. Interview candidates were identified by the first author collaborating with Blekinge Business Incubator (BBI) in Karlskrona, Sweden. The first interview was with a business coach in the incubator, who provided a list of founders of independent (indie) and internal video game startups operating inside larger companies. The interviews were held and recorded in a video conferencing system (Microsoft Teams), each lasting between 60 and 90 minutes. The profiles of the interviewees are shown in Table \ref{tab:case_interviews}. The audio recordings were transcribed and analysed using thematic analysis~\cite{Cruzes2011}. The transcripts were sent back to the interviewees for follow-up questions and clarification.

\begin{table}[ht]
    \centering
    \caption{Overview of interviewees}
    \label{tab:case_interviews}
    \begin{tabular}{clccllcl}
    \hline
       ID & \makecell[l]{Company\\name} & \makecell[c]{Company\\age (years)} & \makecell{Number\\of people} & \makecell[l]{Type of\\company} & \makecell[l]{Game\\ genre} & \makecell[l]{Experience\\(years)} & Role \\ \hline
        A & BBI & & - & Incubator & - & 10 & Business coach \\
        B & \makecell[l]{Mana\\Brigade} & 3 & 5 &  \makecell[l]{Indie\\startup} & Adventure & 3 & \makecell[l]{Founder/\\ CEO} \\
        C & The Station & 10 & 35-40 & \makecell[l]{Internal\\ startup} & Simulation & 14 & \makecell[l]{Founder/\\ Game director} \\
        D & \makecell[l]{Blackdrop\\Interactive} & 8 & 4-5 & \makecell[l]{Internal\\startup} & Warfare & 8 & Founder \\
        \hline
    \end{tabular}
\end{table}

\section{Results}
\label{sec:results}

This section reports our findings by describing six experimentation scenarios. All quotes and information herein are derived from the interview transcripts. 

\subsection{Technical or digital prototyping}
Our interviews reveal that, in the early stages, the main challenge of game development lies not in the ideation process but in the execution and making the game work. Hence, the first purpose of experimentation is to assess the technical feasibility of the team to develop the game. The game's initial idea is usually outlined in a game design document and describes the game at a high level from the user's perspective. The team builds prototypes using a 3D engine, e.g., Unity, to test the game's complexity and scope. In Mana Brigade, a slightly different approach was taken. This company started out performing experiments with a marching cube algorithm\footnote{Marching cubes is an algorithm to extract a 2D surface mesh from a 3D volume.}. This algorithm was then implemented in Unity, and the user experience was tested using VR devices. 

All interviewees agree that technical experimentation is crucial to evaluate their capability to build the game. For example, if they can solve all problems to build a game or need key people with certain skills and expertise. Technical experimentation also showcases their capabilities to potential investors or publishers.

\subsection{Controlled game tests}
Game startups also experiment with external stakeholders, such as end users or players, to evaluate whether they understand the game's concepts and mechanics. In the case of The Station, they hired external game companies to test their game: \textit{ ``[The external video game companies] bring in players. We have a questionnaire that we want them to answer that they rate the game [like] `Was there anything unclear? What did you not like? What did you like?' '' (Interviewee C)}

\subsection{Mock reviews}
In the case of The Station, they asked game journalists to write a mock review and to give a score of their game compared to other games in the same genre. The score was used as an early indicator of what could happen when the game was released. In the case of Mana Brigade, they mentioned that it does not use this approach due to a lack of funding.
 
\subsection{Presenting and pitching in game conferences}
Presenting and pitching new games in video game conferences is a good opportunity to validate assumptions about the game, e.g., the basic idea and its potential market. In these events, video game startups can meet and talk to publishers, investors, or game scouts to get investments from them to build the game. Mana Brigade's first experimentation with external stakeholders was competing in a game competition in 2021. \textit{``For the first iteration, we want it to be multiplayer, and [we want] to explore dungeons. It's like awesome, like real-time events. [But] we got feedback from the [judges] `This doesn't make sense.' So we took that year to iterate on it, and then we wanted to do like it was still single player, but it was still crafting and then adventuring.'' (Interviewee B)}

However, explaining and convincing the game concepts and design to publishers is a big problem. Video game startups need to find ways to explain their game and, at the same time, to find the right publishers.\textit{``[Publishers] get bombarded with hundreds of game ideas they must go through to find that one good game... One publisher publisher wants a game design document, not a PowerPoint. They don't care about the pictures, [while others] want many. It's very hard to know what they want.'' (Interviewee B)} 

\subsection{Social media engagement}
The interviewees expressed that they could use social media platforms, i.e., YouTube or Instagram, to experiment and gain user feedback. For example, by releasing screenshots, images, videos or tutorials on social media and measuring gamers' reactions to these. However, this may not work for indie game startups. They must balance the effort and resources between developing the game and actively maintaining communication with the community and the users.

\subsection{Early release of vertical slice}
Releasing a vertical slice\footnote{A vertical slice is a fully playable portion of a game that shows its developer's intended player experience.} on video game platforms like Steam for user testing may allow game startups to build a player base. It may also give them some small revenue to improve the game, but it could harm their reputation. Besides that, they need to find the right audience for their games. \textit{``The game industry is so big... maybe 100 [new games are published] every day on Steam. It's hard to reach and find your audience and see your game. There is so much information [on Steam], and many games [can easily] get drowned.'' (Interviewee A)}

\section{Discussions and Conclusions}
\label{sec:discussion}

Table~\ref{tab:practices} summarises the six practices we identified and their associated challenges. Some of the practices are present in other contexts, e.g. prototyping. Some are adapted to the context of games, e.g controlled game tests and early release, while some are specific to the game industry, e.g. mock reviews by journalists and presentations in game conferences.

\begin{table}[htbp]
\centering
\caption{Experiment practices and challenges in video game startups}
\label{tab:practices}
\begin{tabular}{p{.25\textwidth}p{.35\textwidth}p{.35\textwidth}}
    \hline
Practice & Purpose & Challenges \\ \hline
Technical/Digital Prototyping & Understanding the game complexity and team capability & Missing skill-sets and expertise in-house \\
Controlled game test  & Understanding if users understand the game concepts and mechanics & Funding to hire professional game testers\\
Early (vertical) release & Build user base and get early revenue & Find the right audience, maintain the reputation \\
Social media engagement & Build user base & Need high effort \\
Mock reviews by game journalists  & Estimate the review score & Funding to hire professional game journalist \\
Presenting and pitching the game & Understanding the market potential and securing funding & Explaining the game's concepts and design to publishers \\ \hline
\end{tabular}
\end{table}

The identified challenges can be related to the experimentation inhibitors experimentation identified by Melegati et al.~\cite{melegati20}. Missing skill sets and expertise and lack of funding to hire game testers or journalists relate to the scarcity of technical and development resources. The need for early releases is associated with time pressure and over-focus on customer base growth in the early phase. However, the difficulty of explaining a game's concepts to publishers might be considered a specific challenge of video game startups. It could be classified as an inhibitor to a valid experiment, as described by Melegati et al. In summary, our study describes the particularities of video game startups and provides evidence to support an existing model in the literature.

This study poses a first step to understanding experimentation within gaming startups. Next, additional video game startups will be studied to further expand on their experimentation practices. We will also expand beyond studying startups that develop games for specific platforms, such as consoles and VR, including other platforms, such as smartphones and tablets. By contrasting and comparing the results, we can improve the generalisability of the findings. Future research could also investigate gaming startups' use of novel technologies, such as artificial intelligence and how these affect their experimentation.

\subsubsection{Acknowledgement}
This work has been supported by ELLIIT; the Swedish Strategic Research Area in IT and Mobile Communications.

%
%
%
\bibliographystyle{splncs04}
\bibliography{reference}

\end{document}